\documentclass[iop, twocolumn, tighten]{emulateapj}
\usepackage{amsmath}
\usepackage{color}
\definecolor{grey}{rgb}{0.5,0.5,0.5}
\definecolor{red}{rgb}{0.8,0.0,0.0}
\newcommand{\thetav}{{\boldsymbol \theta}}
\newcommand{\Thetavf}{{\boldsymbol \Theta}}
\newcommand{\xv}{{\boldsymbol x}}
\newcommand{\sv}{{\boldsymbol s}}
\newcommand{\Tf}{{\mathcal{T}}}

\slugcomment{}
\shortauthors{Kawahara et al.}
\shorttitle{Spectroscopic Coronagraph for PRV}
\begin{document}
\title{Spectroscopic Coronagraphy for Planetary Radial Velocimetry of Exoplanets}

\author{Hajime Kawahara\altaffilmark{1}, Naoshi Murakami\altaffilmark{2}, Taro Matsuo\altaffilmark{3}, and Takayuki Kotani\altaffilmark{4}} 
\altaffiltext{1}{Department of Earth and Planetary Science, The University of Tokyo, 
Tokyo 113-0033, Japan}
\altaffiltext{2}{Division of Applied Physics, Faculty of Engineering, Hokkaido University, Sapporo, Hokkaido, 060-8628, Japan}
\altaffiltext{3}{Department of Astronomy, Faculty of Science, Kyoto University,  606-8502, Japan}
\altaffiltext{4}{National Astronomical Observatory of Japan, 2-21-1 Osawa, Mitaka, Tokyo 181-8588, Japan}
\email{kawahara@eps.s.u-tokyo.ac.jp}

\begin{abstract}
  We propose the application of coronagraphic techniques to the spectroscopic direct detection of exoplanets via the Doppler shift of planetary molecular lines. Even for an unresolved close-in planetary system, we show that the combination of a visible nuller and an extreme adaptive optics system can reduce the photon noise of a main star and increase the total signal-to-noise ratio (S/N) of the molecular absorption of the exoplanetary atmosphere: it works as a spectroscopic coronagraph. Assuming a 30 m telescope, we demonstrate the benefit of these high-contrast instruments for nearby close-in planets that mimic 55 Cnc b ($0.6 \lambda/D$ of the angular separation in the K band). We find that the tip-tilt error is the most crucial factor; however, low-order speckles also contribute to the noise. Assuming relatively conservative estimates for future wavefront control techniques, the spectroscopic coronagraph can increase the contrast to $ \sim 50-130$ times and enable us to obtain $\sim 3-6 $ times larger S/N for warm Jupiters and Neptunes at 10 pc those without it. If the tip-tilt error can be reduced to $\lesssim 0.3$ mas (rms), it gains  $\sim 10-30$ times larger S/N and enables us to detect warm super-Earths with an extremely large telescope. This paper demonstrates the concept of spectroscopic coronagraphy for future spectroscopic direct detection. Further studies of the selection of coronagraphs and tip-tilt sensors will extend the range of application of the spectroscopic direct detection beyond the photon collecting area limit.
\end{abstract}
\section{Introduction}
To date, a small number of exoplanet atmospheres have been characterized via low-resolution spectroscopy from primary and secondary transits and direct imaging. Recently a novel technique to detect exoplanetary dayside atmosphere was successfully accomplished with high-dispersion spectroscopy \citep[][see also \cite{2010Natur.465.1049S} for transmission with a similar technique]{brogi,2012ApJ...753L..25R,2013A&A...554A..82D, 2013MNRAS.432.1980R,2013MNRAS.tmpL.152B,2014ApJ...783L..29L}. In this technique, high-dispersion ($R \sim 100,000$) spectra of the exoplanet + star are analyzed to detect the planet signature. Computing the cross-correlation function between the spectra and a template of molecular lines as a function of the Doppler shifts, allows separation of the planet signature from the stellar absorption lines and the telluric lines. Hence, it detects the change in the radial component of the planet's orbital motion, i.e., planetary radial velocity (PRV). The technique provides precious information on atmospheric compositions. Thus far, the 3-5 $\sigma$ detections of carbon monoxide and water of nearby hot Jupiters have been reported using this technique with CRIRES/VLT and NIRSPEC/Keck. \citet{2014A&A...561A.150D} investigated the cross-correlation signals for molecules not yet detected, methane, acetylene, and hydrogen cyanide.  The application of this technique to reflected light was also proposed by \cite{2013MNRAS.436.1215M}, assuming extremely large telescopes (ELTs). Furthermore, such precise measurements of the PRV will also provide complementary physics, for instance, planetary rotation \citep{kawahara12}.

The targets of the spectroscopic direct detection are currently limited to planets with low star-planet contrast, i.e., hot Jupiters with the equilibrium temperature $T_\mathrm{eff} \gtrsim 1000$ K. An open question is how to improve the signal-to-noise ratio (S/N) of the spectroscopic direct detection in order to apply this technique to other types of planets beyond hot Jupiters. To use larger telescopes is, of course, one of the solutions. ELTs can simply provide an S/N that is $\sim$ 3 times larger than that of current observations. Improvement in efficiency of high-dispersion instruments will also increase the S/N. Beyond the collecting power, if the photon noise from the host star can be reduced, the high-dispersion observations of the exoplanets will be of wider application. High-contrast instruments for direct imaging are designed exactly for the reduction of stellar flux. The aim of this paper is to consider the possibility of spectroscopic corongraphy, that is, an application of the high-contrast instruments to the spectroscopic direct detection technique. 

It is extremely difficult for the coronagraph to be capable of separating the star and the planet near 1 $\lambda/D$, corresponding to 0.1-0.2 AU at 10 pc for the ELTs. However, even if the coronagraph does not reduce the stellar halo at the planet location, but the full contrast difference even a slight reduction benefits this technique. In this paper, we show the visible nuller, as an example of coronagraphs, with the ELTs significantly reducing the relative noise of the planetary radial velocimetry of the dayside emission if the extreme adoptive optics (ExAO) works well. 

The paper is organized as follows. We first summarize the visible nuller and define the quantities that describe its benefit for this use in Section 2.1. We derive the analytical expressions of these quantities as a function of the finite effect of the stellar radius and the tip-tilt contamination in Sections 2.2-2.3. In Section 3, we perform numerical simulations assuming a 30 m telescope with a future ExAO system and consider the speckle noise. We discuss the feasibility of the spectroscopic coronagraph for nearby exoplanets. In Section 4, we also consider ways to further increase the S/N for future studies, as well as the limitations of our techniques. Finally we summarize our results in Section 5.

\section{Analytical Estimate of Gain of the Spectroscopic Coronagraph for Visible Nuller}

In this section, we analytically estimate the gain of the spectroscopic coronagraph. The effects of the stellar angular radius and tip-tilt errors are included in the analytical expressions of the gain in the S/N. We use these expressions to analyze the simulation results in Section 4.

Throughout this paper, we utilize the visible nuller as the coronagraph among various types of coronagraphs. Although we do not claim that the nuller is the best choice for the spectroscopic coronagraph, there are several reasons to use the visible nuller to demonstrate the concept: its simple analytical formulation, small inner working angle (IWA), and compatibility with segmented mirrors for ELTs.

\subsection{Visible Nuller}

The visible nuller, or the visible nulling coronagraph, is the nulling interferometer for a single telescope \citep[e.g.][]{2003SPIE.4860...32M,2004SPIE.5487.1296S,murakami}, which divides a single aperture into two or more apertures, inserts $\pi$ relative phase shifts, and combines the beams. These processes cancel out an on-axis source (i.e. a host star) by a destructive interference and leaves an off-axis source (a planet) unnulled. Here, we briefly summarize several important features of the nuller. Throughout this paper, we consider the simplest two-beam nuller. The detailed description of the visible two-beam nuller is given in the Appendix. 

\begin{figure}[tbh]
  \includegraphics[width=\linewidth]{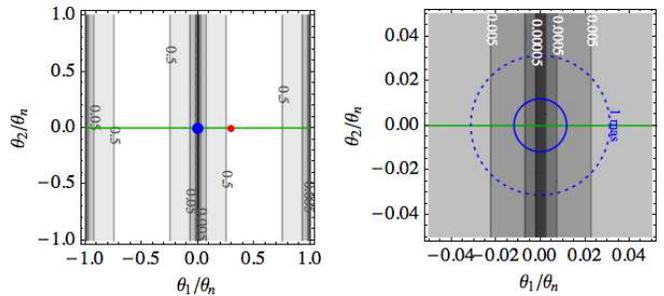}
\caption{Transmittance of intensity for the two-beam nuller and configuration of the exoplanetary system (left). The red and blue points indicate configuration  of the representative warm Jupiter 55 Cnc b ($a=0.11$ AU and $d=12.3$ pc) assuming $\lambda=2.3 \mu$m and $s=15$ m for a 30 m telescope. The right panel displays an enlarged view of the left one around the star. Solid and dotted circles show the stellar disk of the mock 55 Cnc and a 1 mas circle as the typical tip-tilt error, respectively.  \label{fig:transm} }
\end{figure}

The intensity transmittance, $ \Tf(\thetav) $, on the sky plane $\thetav$, for the two-beam nuller is displayed in Figure \ref{fig:transm}. The $x$ and $y$-axis are position angles normalized by the angular length of a fringe, 
\begin{eqnarray}
\theta_n \equiv \frac{\lambda}{s} \approx 32 \, \left( \frac{\lambda}{2.3 \mu \mathrm{m}} \right) \left( \frac{s}{15 \mathrm{m}} \right)^{-1} \mathrm{\,(mas)},
\end{eqnarray}
where $s$ is the amount of shear. The transmittance varies from 0 \% on axis to 100 \% for $\theta_x = \theta_n/2$ along the $x$ direction. Hence, $\theta_n$ can be regarded as a characteristic angle of the two-beam nuller. We assume that a planet is on the $x$ axis of the nuller. This assumption is the most optimistic case when the azimuth angle of the planet is unknown. In general, one needs to determine the roll angle with the strongest signal by observing several angles, or one can utilize the astrometric detection of the planet to determine the adequate roll angle.

We define the normalized separation of the planet by
\begin{eqnarray}
\eta \equiv \theta_p/\theta_n &\approx& 0.31 \, \left( \frac{\theta_p}{10 \mathrm{mas}} \right) \left( \frac{\lambda}{2.3 \mu \mathrm{m}} \right)^{-1} \left( \frac{s}{15 \mathrm{m}} \right),
\end{eqnarray}
where $\theta_p$ is the angular separation of the host star and the planet. We also define the ratio of $\theta_p$ to the stellar angular radius, $\theta_s$, which quantifies the easiness to suppress the finite effect of the stellar radius, by
\begin{eqnarray}
\alpha \equiv \theta_p/\theta_s \approx  20 \left( \frac{a}{0.1 \mathrm{AU}} \right) \left( \frac{R_s}{R_\odot} \right)^{-1},
\end{eqnarray}
where $R_s$ is the stellar radius and $a$ is a star-planet separation. We can relate the radiative equilibrium temperature of the planet to $\alpha$ as 
\begin{eqnarray}
\label{eq:equibtemp}
T_\mathrm{eff} &=& \left( \frac{1-A}{4 \beta_m} \right)^{\frac{1}{4}} \sqrt{\frac{R_s}{a}} T_s \\
\label{eq:aprrox}
&\approx& (0.6-0.8) \, \alpha^{-1/2} T_s
\end{eqnarray}
where $T_s$ is the stellar effective temperature, $A$ is the Bond albedo, and $\beta_m$ is the efficiency of the heat transport (usually $1/2 \le \beta_m \le 1$). Hot Jupiters with $T_\mathrm{eff} = 1500$ K correspond to $\alpha \sim 10$ for G-type stars and warm Jupiters with $T_\mathrm{eff} = 1000$ K have $\alpha \sim 25 $ for G-type stars.

The S/N for the shot noise limit is given by 
\begin{eqnarray}
\label{eq:snma}
\mathrm{(S/N)} = \frac{n(S)}{\sqrt{n(N) + n(S)}} \approx \frac{n(S)}{\sqrt{n(N)}},
\end{eqnarray}
where $n(S)$ and $n(N)$ are the photo-electron counts of the signal and noise. For the dayside PRV, $n(S)=n_p \times \mathrm{EW}/\Delta \lambda$ and $n(N)=n_s$, where $n_s$ and $n_p$ are photo-electron counts from the star and the planet within $\Delta \lambda$ and $\mathrm{EW}$ is the equivalent width of the absorptions. 

We denote the change of throughput for the star and the planet after adding the nuller by $\rho_s$ and $\rho_p$ (the nuller works as the stellar flux $F_s \to \rho_s F_s$ and the planet flux $F_p \to \rho_p F_p$). Then we can write $\rho_s$ and $\rho_p$ by the products of the transmittance function $\Tf$ and the instrumental throughput of the nuller $\delta(q)$,
\begin{eqnarray}
\rho_s &=& \delta(q) \Tf_s \\
\rho_p &=& \delta(q) \Tf_p,
\end{eqnarray}
where $\Tf_s$ and $\Tf_p$ are the intensity transmittance function at the positions of the host star and the planet, respectively. We assume that the total throughput of the nuller is equivalent to the deficit of aperture, corresponding to the ratio between the overlapped area of the beams and the effective area of the telescope (see Eq. [\ref{eq:tra}] ). Hence, $\delta(q)$ depends on the ratio of the telescope diameter and the shear $q \equiv s/D$. We ignore other factors in the nuller system that may decrease the throughput, such as the throughput of the optical system.

From Equation (\ref{eq:snma}), the nuller increases (or decreases) the S/N of the PRV as
\begin{eqnarray}
\mathrm{(S/N)} \to (\rho_p/\sqrt{\rho_s}) \times \mathrm{(S/N)} = M \times \mathrm{(S/N)}, 
\end{eqnarray}
where we quantify the benefit to use the nullers by the gain $M$ in the S/N as 
\begin{eqnarray}
\label{eq:meritgain}
M &\equiv& \psi \sqrt{\delta(q)} \\
\label{eq:psi}
\psi &\equiv& \Tf_p/\sqrt{\Tf_s},
\end{eqnarray}
where $\psi$ depends on the transmittance pattern of the system of the nuller, the configuration of the star and the planet on the sky plane, and the stellar radius. 

The nuller changes the apparent magnitude of a star $m_\star$ to
 an {\it effective} apparent magnitude as,
\begin{eqnarray}
m_\mathrm{eff} = m_\star - 2.5 \log_{10} (\delta(q) {\Tf_s}).
\end{eqnarray}
We define an {\it effective} star-planet contrast by,
\begin{eqnarray}
C_\mathrm{eff} = \frac{\Tf_p}{\Tf_s} C_\mathrm{sp}.
\end{eqnarray}

\subsection{Finite Stellar Radius}

If the phase aberration is negligible, the transmittances of photons from the star and planet are given by
\begin{eqnarray}
\Tf_s &=& \overline{\Tf(0)} \\
\Tf_p &\sim& \Tf({\thetav}_p),
\end{eqnarray}
where $\overline{\Tf}$ takes an average over the stellar disk and ${\thetav}_p$ is the angular vector of the planet. We ignore finiteness of the planetary disk. With the above equations, we can rewrite Equation (\ref{eq:psi}) as
\begin{eqnarray}
\psi &=&
\frac{ \Tf(\theta_p)}{\sqrt{\int_{-\theta_s}^{\theta_s} d \theta \,\, 2 \sqrt{\theta_s^2 - \theta^2} \Tf(\theta)/\pi \theta_s^2 }}  \nonumber \\
&=& \sin^2{(\pi \eta)} \left(\frac{1}{2} - \frac{J_1(2 \pi \eta/\alpha)}{2 \pi \eta/\alpha} \right)^{-1/2}, \\
\label{eq:app2}
&\approx&  \frac{2 \alpha \sin^2{(\pi \eta)}}{\pi \eta} \,\, \mbox{for $\alpha \gg \eta$} 
\end{eqnarray}
where $J_1 (x)$ is the Bessel function of the first kind.  


\subsection{Tip-tilt}

An angular vector of the instance tip-tilt error, $\thetav_t$ changes the transmittances of the star and planet as
\begin{eqnarray}
\Tf_s &=& \overline{\Tf (\thetav_t)} \\
\Tf_p &\sim& \Tf ({\thetav}_p + \thetav_t).
\end{eqnarray}
For the two-beam nuller, the tip-tilt error in the $x$ direction ($\theta_{t,x}$) can change the transmittance. The tip-tilt error in the $y$ direction does not affect it. Then, we obtain 
\begin{eqnarray}
\psi = \displaystyle{\frac{\sin^2{[\pi (\eta \pm \tau)]} }{\sqrt{ \frac{1}{2} - \cos{(2 \pi \tau)} \frac{J_1(2 \pi \eta/\alpha)}{2 \pi \eta/\alpha} }}}, 
\label{eq:app2tts}
\end{eqnarray}
where the $\pm$ sign corresponds to the direction of the tip-tilt to the planet and we define the relative tip-tilt error, 
\begin{eqnarray}
\tau \equiv \frac{|\theta_{t,x}|}{\theta_n} \approx 0.03 \, \left( \frac{|\theta_{t,x}|}{1 \mathrm{mas}} \right) \left( \frac{\lambda}{2.3 \mu \mathrm{m}} \right) \left( \frac{s}{15 \mathrm{m}} \right)^{-1}.
\label{eq:tau}
\end{eqnarray}
Assuming that $\tau \ll \eta$, we use 
\begin{eqnarray}
\psi \sim \displaystyle{\frac{\sin^2{[\pi \eta]} }{\sqrt{ \frac{1}{2} - \cos{(2 \pi \tau)} \frac{J_1(2 \pi \eta/\alpha)}{2 \pi \eta/\alpha} }}}, 
\label{eq:app2tt}
\end{eqnarray}
to compare with the result of simulations in Section 3. Let us consider the case that the tip-tilt error dominates over the stellar angular radius.
Taking the limit $\alpha \to \infty $, we simply obtain 
\begin{eqnarray}
\psi_\mathrm{lim} (\theta_p,\theta_{t,x}) \equiv \frac{\sin^2{(\pi \eta)}}{\sin{(\pi \tau)}} = \frac{\Tf(\theta_p)}{\sqrt{\Tf(\theta_{t,x})}}.
\label{eq:app2ttslim}
\end{eqnarray}
We find that the difference between $\psi$ and $\psi_\mathrm{lim}$ is within $\sim 10$ \% if $\theta_{t,x}>\theta_{s}$, equivalently $\tau > \eta/\alpha$. One may intuitively understand this situation through the right panel of Figure \ref{fig:transm}. This panel shows the stellar angular radius (solid circle) for the nearby warm Jupiter system, 55 Cnc b, and our fiducial tip-tilt error 1 mas on the transmittance map.

\begin{figure*}[tb]
\begin{center}
  \includegraphics[width=0.33 \linewidth]{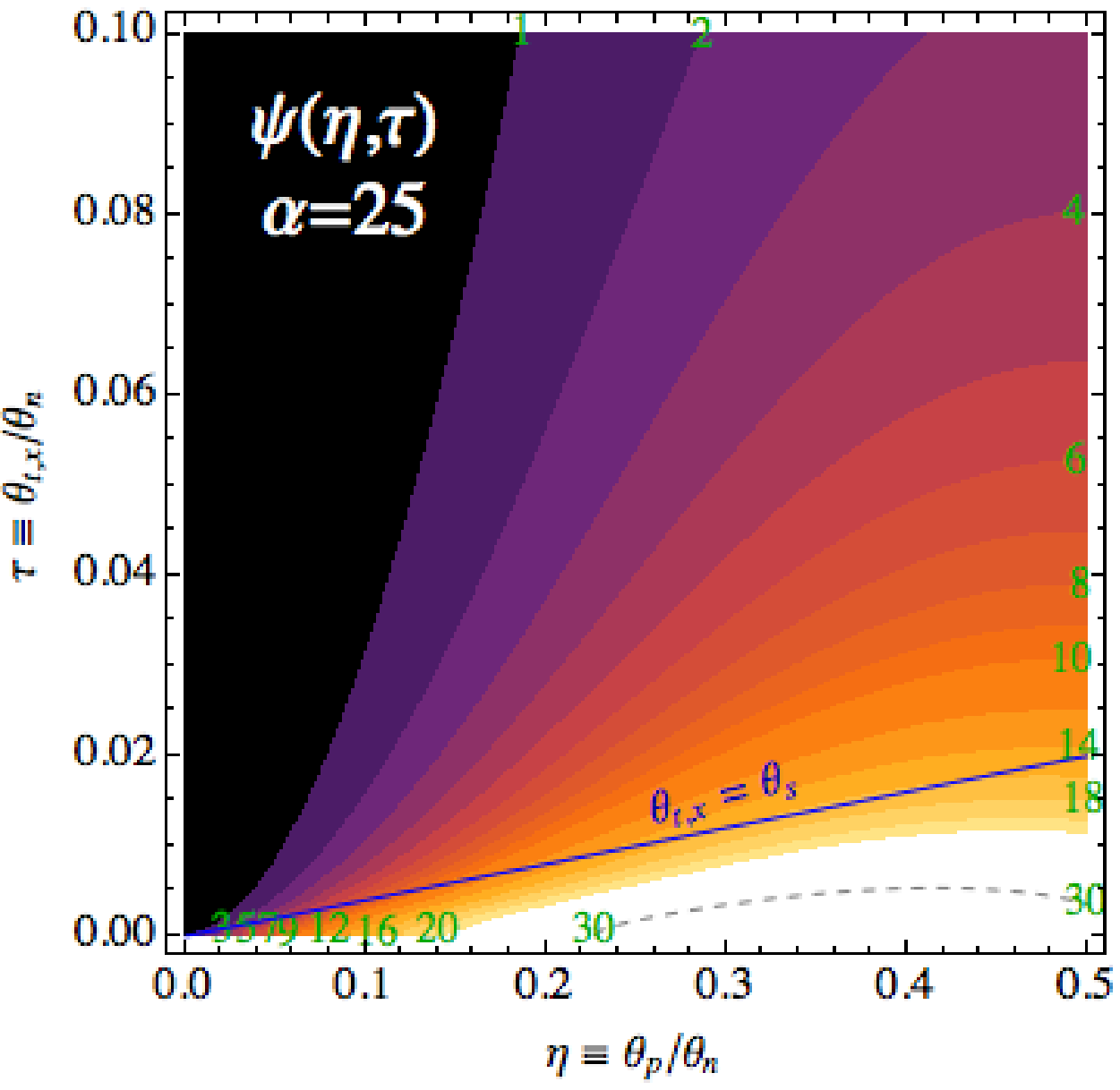}
  \includegraphics[width=0.33 \linewidth]{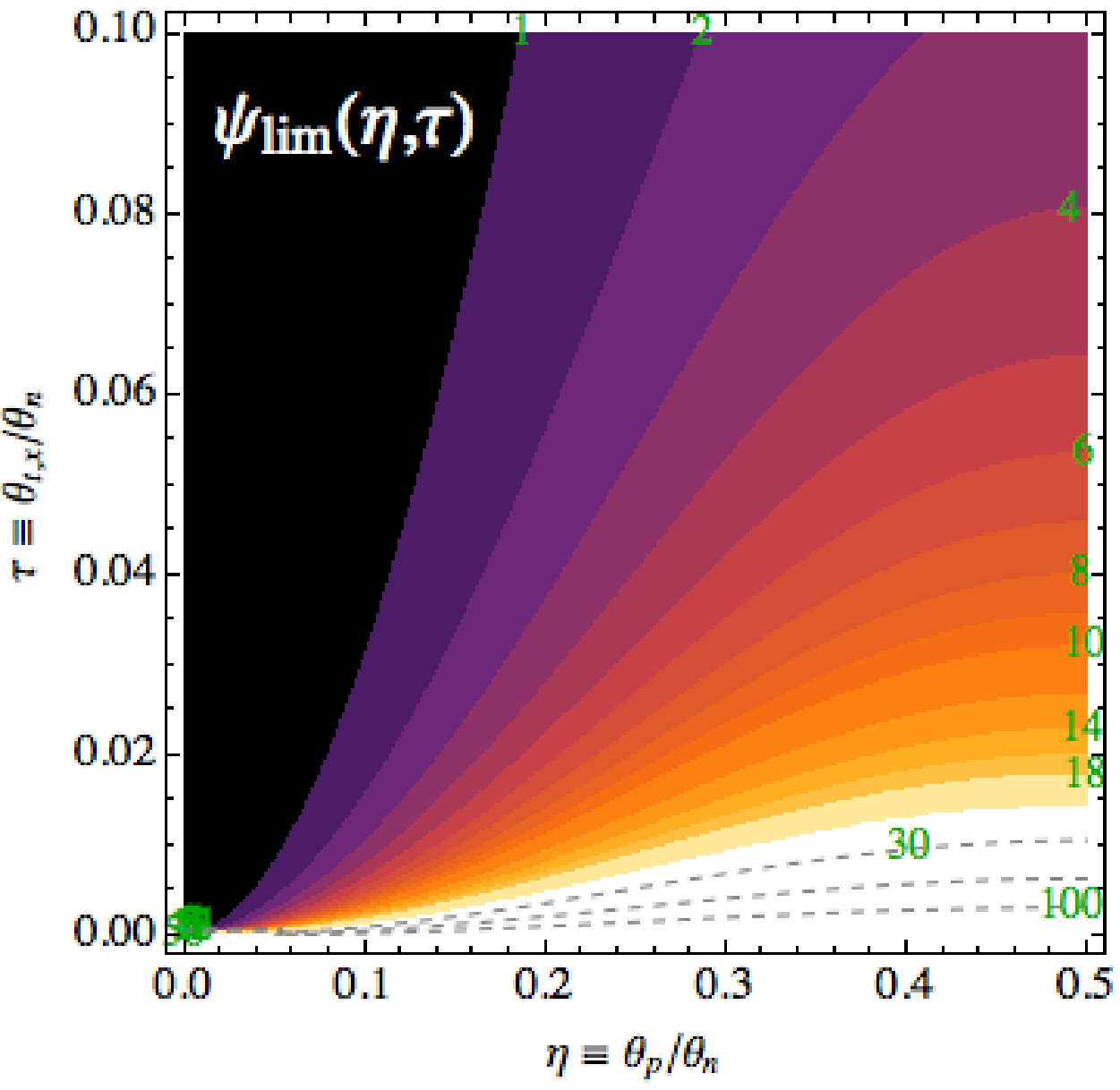}
\end{center}
\caption{$\psi$ for $\alpha=25$ (corresponding to warm Jupiters around G-type stars) and $\psi_\mathrm{lim}$ as a function of the normalized tip-tilt $\tau$ and the normalized angular separation $\eta$. The gain in the S/N is defined by $M = \sqrt{\delta(q)} \psi$. Since $\delta (q)$ is not significantly smaller than unity, $\psi$ dominates the gain.      \label{fig:etatau}}
\end{figure*}

The left panel in Figure \ref{fig:etatau} displays $\psi$ as a function of the planet position, $\tau$, and the tip-tilt error $\eta$ including the effect of the stellar finite radius with $\alpha=25$, and roughly corresponding to the ratio of a G-type stellar radius to semi-major axis of warm Jupiters or Neptunes. Recalling that the gain in the S/N, $M= \sqrt{\delta(q)} \psi$ and $\delta(q) \lesssim 1$, several to several tens of gain in the S/N is expected in the wide range of $\eta$ and $\tau$ from the analytical estimates with the effects of the tip-tilt and the stellar radius. For instance, a typical set of for nearby warm Jupiters for $\theta_{t,x}=1$ mas, $\eta=0.3$, and $\tau=0.03$ yields $\psi \sim 7$.  
  
We also show $\psi_\mathrm{lim}$ in the right panel, which disregards the effect of the stellar radius (Eq. [\ref{eq:app2ttslim}]). As shown in these panels, it makes little difference in the gain if the tip-tilt error is larger than the stellar angular radius (above the blue line in the left panel). However, the gain under the blue line in the left panel is significantly suppressed by the stellar radius. For the extreme case of the very small tip-tilt error, the stellar radius dominates the gain, though it might be premature to consider such a case. Hence, the theoretical maximum gain of the S/N for the extreme case is $M \sim 30$ for the warm Jupiter, Neptune, and super-Earth.

\section{Simulations with Wavefront Errors}
In the previous section, we analytically estimated the gain in S/N considering the tip-tilt error and the stellar finite radius. However, we still have other possible effects. One important noise is the speckle due to the phase aberration. Moreover, we only consider the instant tip-tilt error in the previous section. In reality, the tip-tilt error and the speckle vary from time to time. There are also other minor effects of the finiteness of the sheared telescope, the secondary mirror, the spiders, and the extract region. Since these effects are difficult to be expressed analytically, we perform mock observations to examine and analyze these effects with the aid of the analytical expressions described in Section 2. 

The assumptions of the mock observations are summarized in Table \ref{tab:sim}. We assume a 30 m telescope with a next generation of ExAO for the wavefront errors after the ExAO. We use the K band because a low star-planet contrast is expected and there are a lot of strong molecular lines of carbon monoxide and methane in this band. The carbon monoxide and methane are the most dominant molecules for the thermal chemical equilibrium in gas giants. Mixing ratios of carbon monoxide and methane depend on the temperature and C/O ratio \citep{2012ApJ...758...36M}. For the gas planets $\lesssim 1200$ K, the methane molecular lines are the most promising lines for the spectroscopic detection. We use the band center of methane $\sim 2.3 \mu$m as the representative wavelength.

The simulated wavefront after the ExAO that we use was developed for the simulation of the ExAO, assuming the Thirty Meter Telescope (T. Matsuo et al., in preparation). The procedure for computing the wavefront is (1) adapting the weights and altitudes of the turbulence layers from MASS and Scidar measurements at the top of Maunakea \citep{2005PASP..117..395T} to the atmospheric parameters, (2) including the Fresnel propagation through atmosphere, (3) setting an accurate model of an ideal 128$\times$128 actuator system with linear and non-hysteresis effects as the correcting device, (4) applying a Pyramid wavefront sensor with the fixed position \citep[e.g.,][]{2004SPIE.5490.1177V} to the Wavefront measurement system with a 2 kHz sampling, (5) setting the wavefront sensing wavelength to 0.8 $\mu$m, and (6) observing an I=6 magnitude star with the adaptive optical system \citep[similar to that of 55 Cnc, I=5][]{2003AJ....125..984M}. The simulated wavefront has a $\sim 1$ mas rms tip-tilt vibration and 0.92 of the Strehl ratio. The tip-tilt errors in this simulation are created only by the phase aberration. In reality, the instrumental pointing errors also contribute to the gain, in principle. For the high contrast observations, it should be smaller than the rms of the atmospheric tip-tilt error and we ignore them in this simulation. 

\begin{table}[!htb]
\begin{center}
\caption{Assumptions of the Mock Observations \label{tab:sim}}
  \begin{tabular}{cc}
   \hline\hline
\multicolumn{2}{c}{For computing the gain in the S/N $M$} \\
\hline
Telescope diameter & 30 m \\
Shear & $s=15,12,9$ m \\
Secondary mirror and gaps & Mocked TMT \\
Bands & K (2.3 $\mu$m) \\
\hline\hline
\multicolumn{2}{c}{Additional assumptions for Figure \ref{fig:etametapc}} \\
\hline
Number of lines & $N_\mathrm{CCF}=$ 100 \\
Efficiency of a high-dispersion instrument & $\epsilon=0.1$ \\
Total exposure time & 20 hr \\
\end{tabular}
\end{center}
\end{table}

\begin{figure*}[!htb]
  \includegraphics[width=0.49 \linewidth]{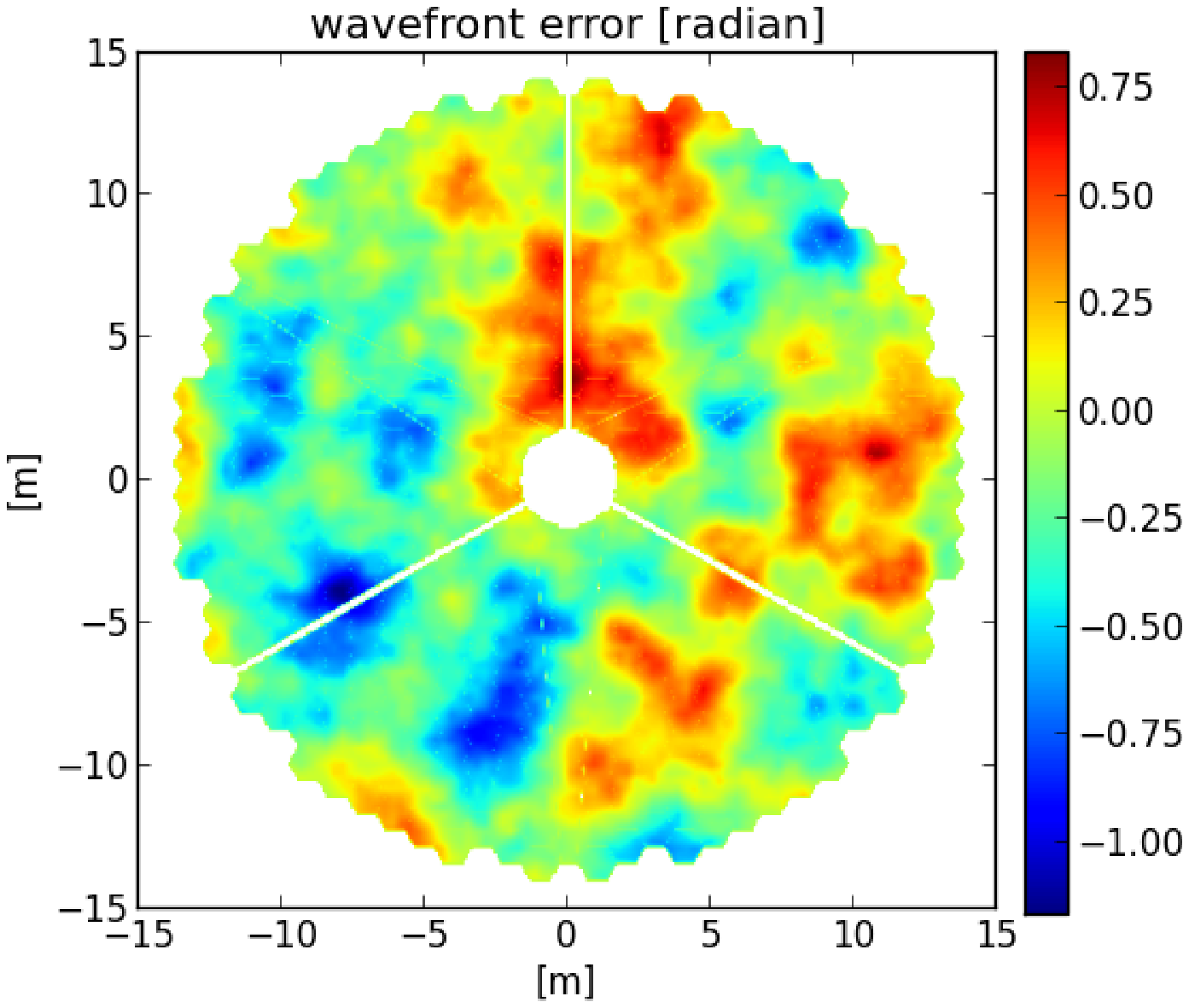}
  \includegraphics[width=0.40 \linewidth]{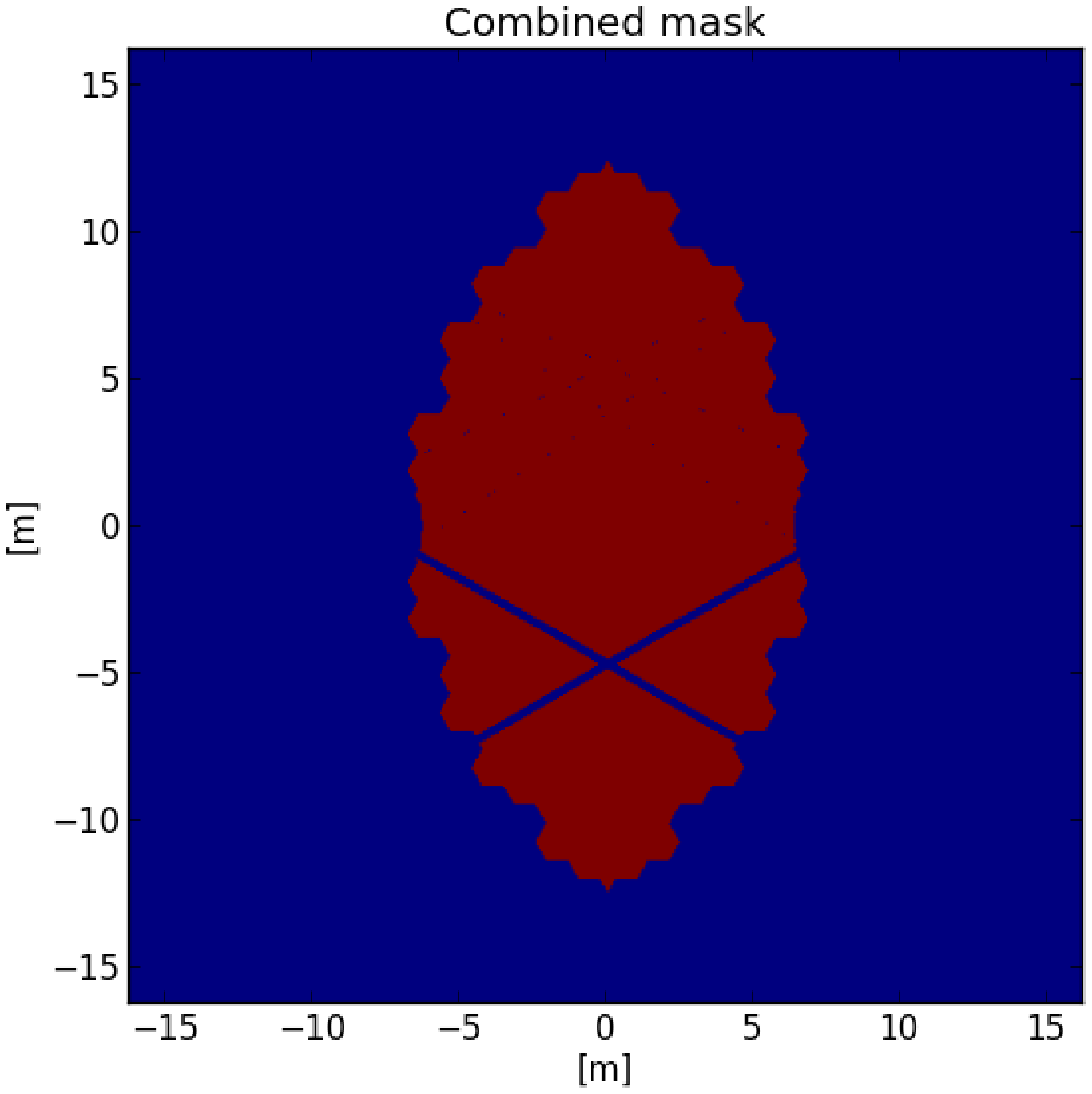}
\caption{Example of a simulated wavefront (K band)  and an overlap region of the two beams for $s=0.5$. \label{fig:wfea}}
\end{figure*}

We compute the complex amplitude of the pupil by sampling 50 points within the stellar disk. The complex amplitude from the point $\theta$ is written as 
\begin{eqnarray}
A_{P,\mathrm{single}} (\xv,\thetav) = M_t (\xv) V (\thetav) e^{2 \pi i \xv \cdot \thetav + i \phi(\xv)} ,
\end{eqnarray}
where  $\phi(\xv)$ is the wavefront error at the pupil (the left panel in Figure \ref{fig:wfea}), $V(\thetav)$ is the amplitude of sky plane at $\thetav$ and $\xv = {\bf X}/\lambda$ (${\bf X}$ is a physical coordinate on the pupil). $M_{t} (\xv)$ is the mask function of a telescope, which includes the boundary of the telescope, the shadow of a secondary mirror and spiders. The complex amplitude of the sheared pupil from $\theta$ is given by  
\begin{eqnarray}
\label{eq:apaaa}
A_P (\xv,\thetav) &=&  M_o(\xv) \frac{1}{n} \sum_k^n W_k A_{P,\mathrm{single}} (\xv-\sv_k,\thetav) \\
 &=& M_o(\xv) \frac{|A_0 (\thetav)|}{n} \sum_k^n W_k e^{2 \pi i (\xv -\sv_k) \cdot \thetav + i \phi(\xv-\sv_k)}, \nonumber \\
\end{eqnarray}
The mask function $M_o (\xv)$ of the overlapped pupils is computed from $M_{t} (\xv)$ (see the Appendix). The shape of $M_o (\xv)$ for $s=15$ m is shown in the right panel of Figure \ref{fig:wfea}. Then, the complex amplitudes of the star and the planet at the focal plane are expressed as  
\begin{eqnarray}
\label{eq:ap}
A_{F,\star} &=& \mathbb{FT}_\xv \left[\int_{|\thetav| \le \theta_s} d \thetav A_P (\xv,\thetav)\right] \\
A_{F,p} &=& \mathbb{FT}_\xv \left[ A_P (\xv,\thetav_p)\right],
\end{eqnarray}
where $\mathbb{FT}_\xv$ is the Fourier transform of $\xv$.  

\begin{figure}[tb]
  \includegraphics[width=0.7\linewidth]{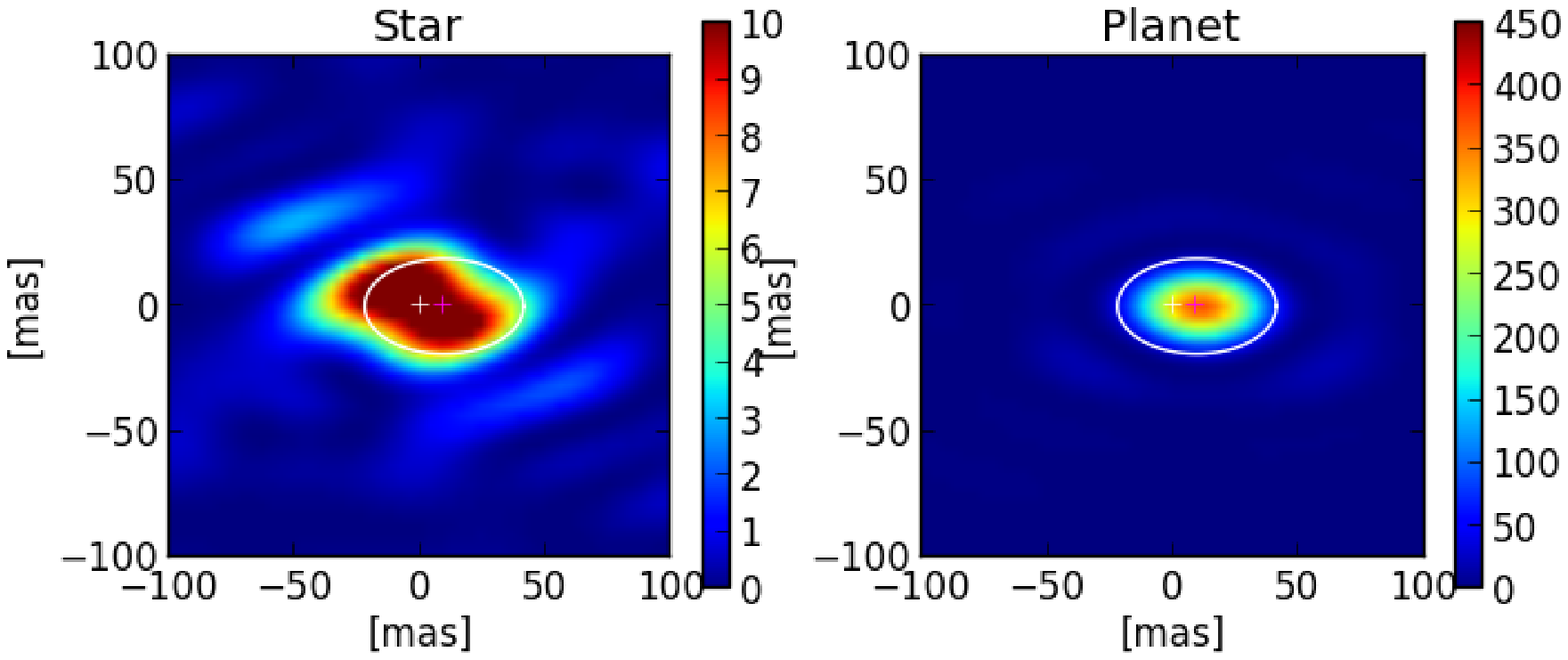}
  \includegraphics[width=0.7\linewidth]{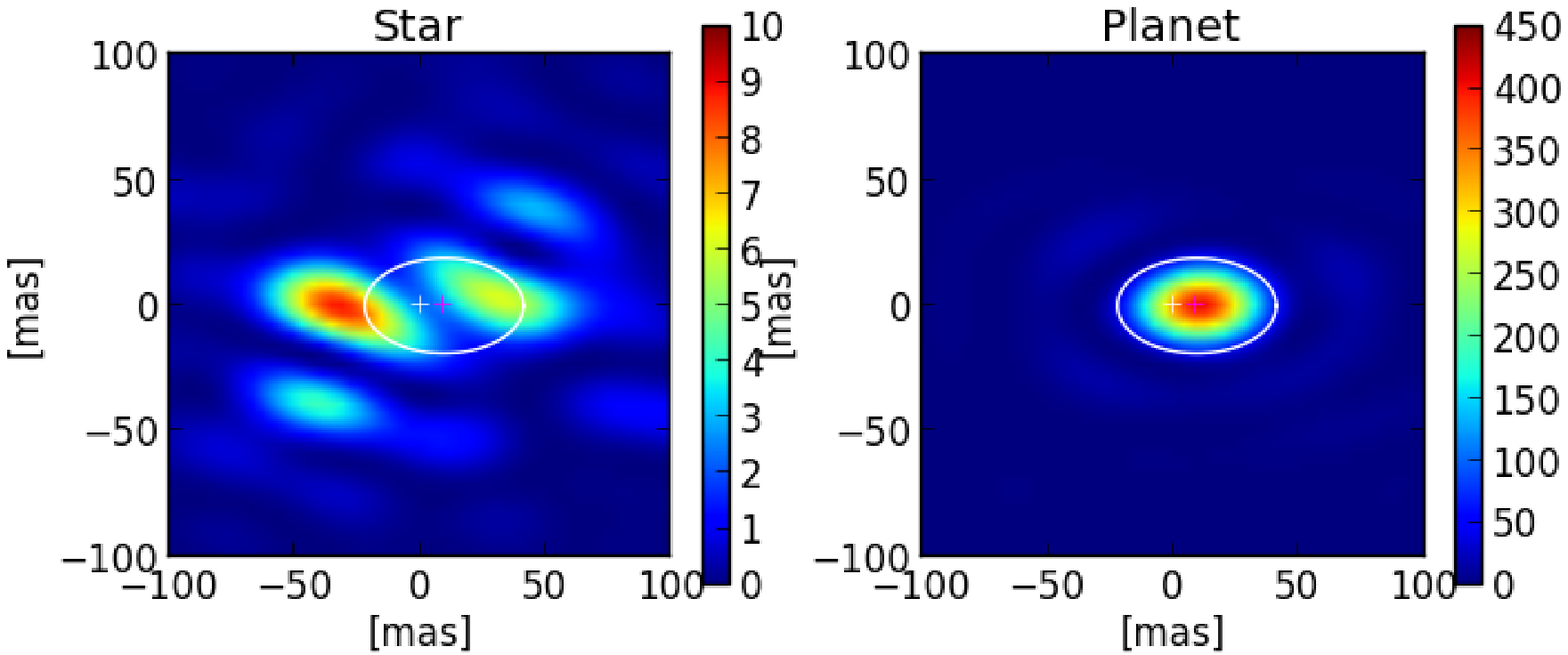}
  \includegraphics[width=0.7\linewidth]{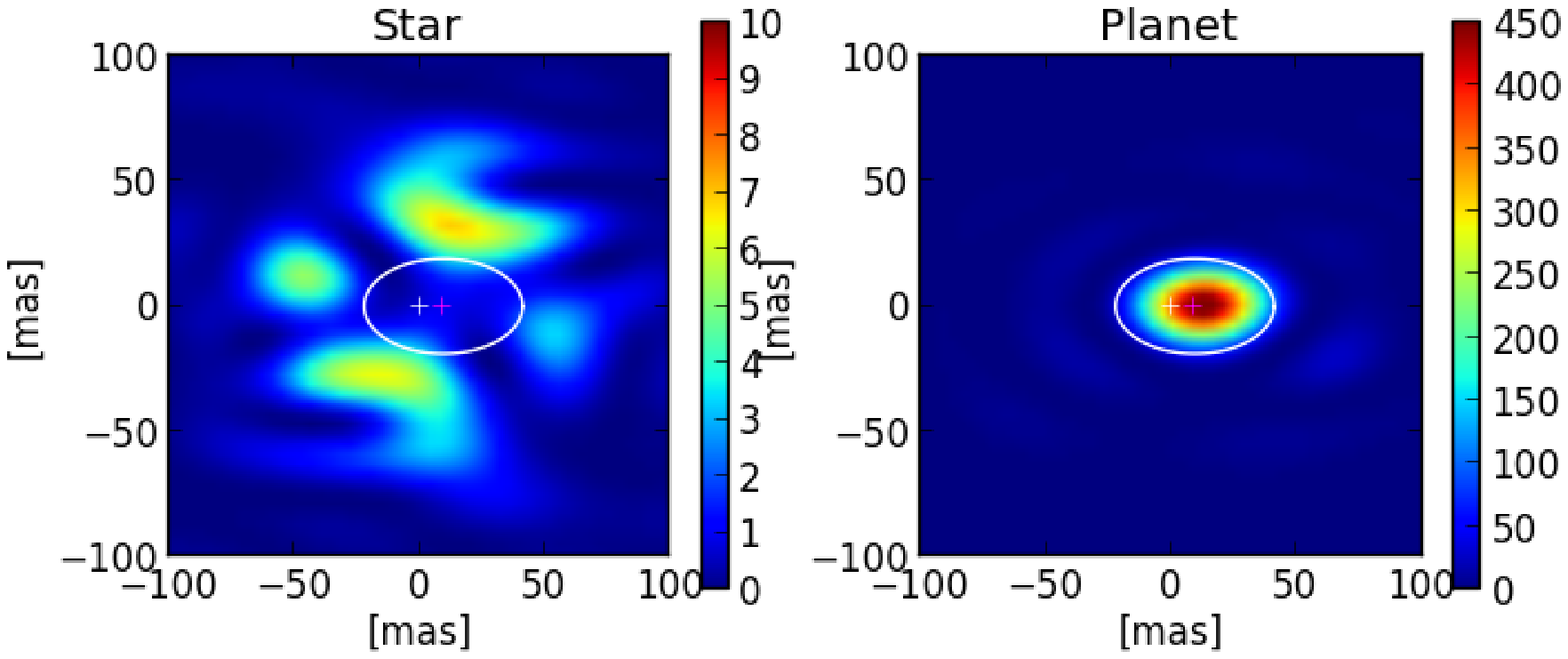}

\caption{Examples of the simulated instant focal plane for the star and the planet mocking the 55 Cnc b system. Top panels: images of the star (left) and the planet (right) for the large instant tip-tilt errors $\theta_{t,x} \sim 1$ mas ($M=1.5$, where $M$ is the gain in the S/N defined by Eq. [\ref{eq:meritgain}]). The position of the star and the planet are indicated by the white and red crosses. The dominant leakage of star light is due to the tip-tilt.  The middle and bottom panels display the images for $\theta_{t,x} \sim 0.5$ mas, but with $M=2.5$ and $5.3$, respectively. In these cases, the dominant leakage is likely to be the speckle noise from higher order contribution of the phase aberration than the tip-tilt because the signal from the star is not in the stellar center. We note that the symmetric patterns are artificial because of the lack of the intensity errors of the wavefront in this simulation. Intensity both for the star and the planet shown by color is normalized by its total flux. 
\label{fig:focal}}
\end{figure}

As a target of the mock observation, we choose a nearby warm Jupiter, 55 Cnc b. A summary of the 55 Cnc b system parameters is listed in Table \ref{tab:ffcnc}. We picked 100 snapshots every 50 ms of the focal planes for the mock 55 Cnc and 55 Cnc b. Examples of snapshots are shown in Figure \ref{fig:focal}. One snapshot has a 1 ms exposure.  We analyze the star and planet lights within the ellipses in Figure \ref{fig:focal} as the noise and signal, respectively. 

\begin{table}[!htb]
\begin{center}
\caption{Parameters of 55 Cnc b \label{tab:ffcnc}}
  \begin{tabular}{ccc}
   \hline\hline
   Parameter & Symbol & Value \\
\hline
Distance & $d$ & 12.3 pc \\
Semi-major axis & $a$ & 0.11 AU \\
Stellar radius & $R_s$ & 0.94 $R_\odot$ \\
K-band magnitude & $m_K$ & 4.0 \\
\hline 
\multicolumn{3}{c}{Normalized values for the K-band, $D=30$m, $s=15$m} \\
\hline 
Angular separation & $\theta_p$ & 9.3 mas \\
Stellar radius & $\theta_s$ & 0.36 mas \\
& $\alpha$  &  25.9 \\
& $\eta$ & 0.29 \\
\hline 
\multicolumn{3}{c}{Simulation results} \\
\hline 
Gain in the S/N & $M$ & 2.8 \\
\end{tabular}
\end{center}
\end{table}

\begin{figure*}[tb]
  \includegraphics[width=\linewidth]{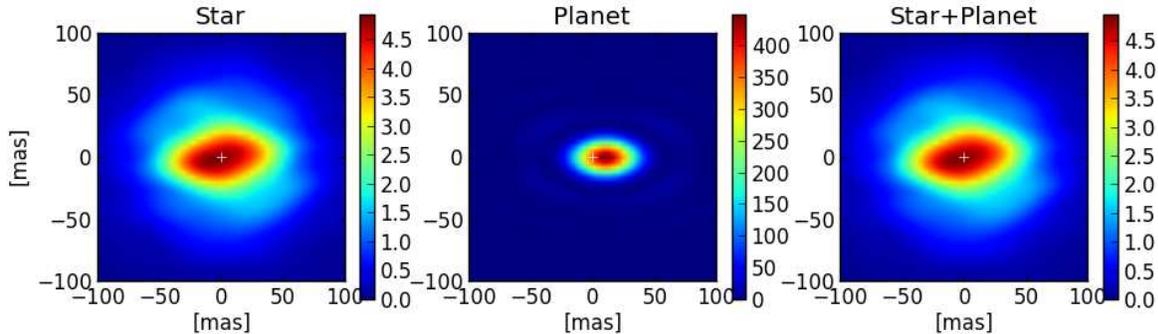}
\caption{Simulated, integrated focal images for the 55 Cnc b system, corresponding  to the star, the planet, and the star+planet system from left to right. The normalization of the left and middle panels are chosen to provide the same intensity when the flux is integrated over the planes. In the right panel, we multiply $10^{-4}$ to the planet image to mock a planet with $C_\mathrm{sp}=10^{-4}$. Scales of intensity both for the star and the planet shown by color are normalized by its total flux and are averaged over 100 snapshots.\label{fig:focalallno}}
\end{figure*}

Figure \ref{fig:focalallno} displays the simulated focal planes assuming 55 Cnc b integrated over 100 snapshots. Because the high-dispersion instruments need long exposures, we estimate the throughput and the gain in the S/N from this ensemble average of the snapshots. As shown in the right panel, even assuming the low star-planet contrast, $10^{-4}$, one cannot identify the planet separately from the star because the planet is within, or very close to, the PSF core. In this sense, this type of observation cannot be regarded as direct imaging. Nevertheless, the contrast increases 51 times larger than the original one ($C_\mathrm{eff}/C_\mathrm{sp} = \Tf_p/\Tf_s=51$).  The simulation for 55 Cnc b exhibits $M = 2.75$, i.e. $\sim$ 3 times larger S/N than that with no coronagraph. Thus, we expect that observations with the ExAO+nuller system on a 30 m telescope will give an order of magnitude ($\times 10$) larger S/N than that obtained by current observations with 10 m telescopes. We note that the planet intensity itself varies by $\sim 10$ \% due to the change of the transmission, however, it does not degrade the gain.

\begin{figure}[htb]
  \includegraphics[width=0.99 \linewidth]{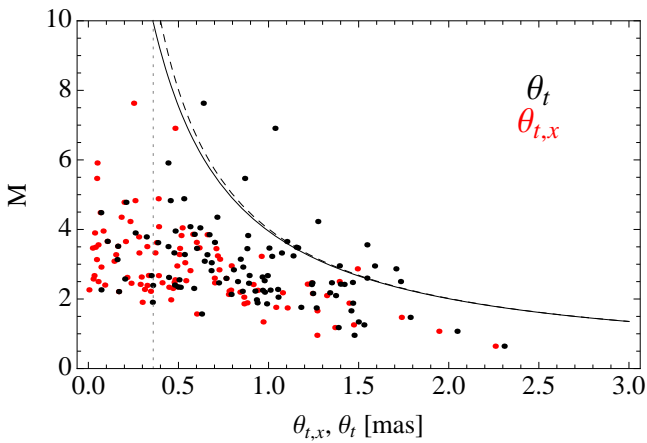}
\caption{Tip-tilt angle for each snapshot $\theta_t$ and tip-tilt in the $x$ direction $\theta_{x,t}$ (red) as a function of the gain, $M$, in the mock observation of 55 Cnc b, assuming a 30 m telescope with the ExAO. The solid curve is the analytical model for tip-tilt error (Eq. [\ref{eq:app2tt}]) with both the stellar radius. Dashed curve is the tip-tilt limit approximation ($M = \psi_\mathrm{lim} \sqrt{\delta(q)}$). A dotted line corresponds to the stellar angular radius, $\theta_s$.\label{fig:tiptilt}}
\end{figure}

The gains with and without including the finiteness of the star in the simulation are $M=2.75$ and $M = 2.73$, respectively. The finite stellar radius effect is negligible. To understand the dominant errors, we also compute $M$ for each snapshot and measure the tip-tilt by a linear fitting of the wavefront on the pupil.  Figure \ref{fig:tiptilt} presents $M$ for each snapshot as a functions of the tip-tilt error for the $x$ direction (red) and the total tip-tilt error (black points). For large $\theta_{t,x}$, $M$ roughly distributes according to the analytical curve for the tip-tilt limit (Eqs. [\ref{eq:app2tt}] and [\ref{eq:app2ttslim}]). The top panel in Figure \ref{fig:focal} shows an example of the focal plane image for large $\theta_{t,x} \sim 1$ mas. For smaller $\theta_{t,x}$, there are relatively large dispersions of $M$ for the same $\theta_{t,x}$. The middle and bottom panels in Figure \ref{fig:focal} show the images of the focal plane for similar $\theta_{t,x} \sim 0.5$ mas. As shown in both panels, the dominant components of stellar leakage are not located at the stellar center. We interpret these components as speckles from the low-order wavefront errors, but higher than the tip-tilt. With the analytical expressions (Eqs. [\ref{eq:app2}] and [\ref{eq:app2tt}]), we obtain $M=22.3$ and $M=5.5$, considering the finite stellar radius effect only and the tip-tilt (adopting the rms value of the $x$ direction of the tip-tilt 0.7 mas) $+$ the stellar radius effect, respectively. Because $M=2.8$ was derived by the full simulation, we conclude that the tip-tilt is the largest component in the stellar leakage, but, the speckles from higher order terms also have non-negligible contributions.
    
Accurate feasibility for the PRV is difficult to compute because one must consider the complex telluric contaminations and systematic noises of the correlation method as done by \citet{2013MNRAS.436.1215M}, and moreover, we do not know the atmospheric composition of the planet. Instead, we estimate the expected S/N by a simple statistical argument \citep[e.g.][]{2013MNRAS.436.1215M},
\begin{eqnarray}
\label{eq:snest}
&\,&\mathrm{S/N} = M \sqrt{N_\mathrm{CCF}} \mathrm{(S/N)_{spectra}} p \, C_\mathrm{sp} \\
&\approx& 8 M \frac{D}{30 \mathrm{m}} \frac{p \, C_\mathrm{sp}}{10^{-5}} \left( \frac{N_\mathrm{CCF}}{100} \frac{F}{F_{-11}} \frac{T_\mathrm{exp}}{20 \mathrm{hr}} \frac{\epsilon}{0.1} \right)^{1/2} \left(\frac{R}{10^5}\right)^{-1/2},  \nonumber \\
\end{eqnarray}
where $N_\mathrm{CCF}$ is the number of molecular lines,  $F_{-11} = 10^{-11} \mathrm{(W/m^2/\mu m)}$ is the stellar flux, corresponding to $m_K=3.8$, $R$ is the spectral resolution, $T_\mathrm{exp}$ is the effective exposure time, and  $\epsilon$ is the efficiency of the high-dispersion instrument, which usually depends on the slit size and the core size of the beam (i.e., the performance of the AO). Because the depth of the lines to the continuum should be shallower than the star-planet contrast, we denote   the line contrast ratio by $p \, C_\mathrm{sp} $, where $p$ is the line depth to the planetary emission \citep[e.g.][]{2013MNRAS.tmpL.152B}. To avoid the confusion, we put $p=1$ and interpret $N_\mathrm{CCF}$ as the effective number of the saturated lines to the planetary emission. If adopting the case of $\sim 6 \sigma$ detection by VLT/CRIRES for tau Boo b by \citet{brogi}, $D=10$m, $T_\mathrm{exp}\sim 18$ hr, $\epsilon=0.01$, \footnote{Instrumentation Division CRIRES User Manual} $N_\mathrm{CCF} \sim 25$, $R = 87000$, $m_K=3.4$,  $p C_\mathrm{sp} = 10^{-4}$ and $M=1$, Equation (\ref{eq:snest}) provides $\mathrm{S/N} \sim 7$. Hence we use Equation (\ref{eq:snest}) to convert $M$ to the expected S/N or the $5 \sigma$ detection limit of the line contrast, $C_{\mathrm{lim}}$.

We note that the typical intensity of the sky background emission is $<10^{-14}$ $(s/10^3$ $\mathrm{mas^2})$ $\mathrm{ergs/s/cm^2/\mu m}$ \footnote{http://www.gemini.edu/sciops/telescopes-and-sites/observing-condition-constraints/ir-background-spectra}, corresponding to $m_K \sim 18$. Because the effective decrease of stellar magnitude by the nuller is $\sim 6$ for our fiducial assumptions of the tip-tilt error, we can safely ignore the sky background emission for the bright nearby stars with the stellar magnitude much smaller than $m_K=12$. 

The star-planet contrast is estimated by
\begin{eqnarray}
C_\mathrm{sp,est} (a) &=& \Phi(\beta) \left( \frac{R_p}{R_s} \right)^2 \frac{e^{hc/\lambda k T_s} -1}{e^{hc/\lambda k T_\mathrm{eff}}-1},
\end{eqnarray}
where we assume that the planet emits black-body radiation with the radiative equilibrium temperature (Eq. [\ref{eq:equibtemp}] with $A=0.3$) and that the heat received from the star is instantly redistributed all over the planet, $\beta_m=1$, or is inefficiently redistributed, $\beta_m=0.5$. The $\Phi(\beta)$ is the phase dependence of the emission. Adopting $\beta=\pi/2$ (position where the angular separation corresponds to the semimajor axis), we simply assume that the contrast is between the two extreme cases: (1) $\Phi(\pi/2)=1$ for $\beta_m=1$ and (2) $\Phi(\pi/2)=1/2$ for $\beta_m=0.5$.

Adopting the parameters for 55 Cnc in Table \ref{tab:ffcnc}, we obtain $C_\mathrm{sp,est} = (3 - 6) \times 10^{-6}$ assuming that the planetary radius is the Jupiter radius, $R_p=R_J$, $A=0.3$ and $\epsilon=0.1$. Though there are a lot of molecular lines of methane and carbon monoxide in the K band, the number of available lines, $N_\mathrm{CCF}$, depends on telluric absorptions of the telescope sight, the size of instruments, and so on. Here we conservatively choose $N_\mathrm{CCF} = 100$, which is larger than, but of the same order of the value in \citet{brogi}. Then, Equation (\ref{eq:snest}) yields a 6-12 $\sigma$ detection for a 20 hr observation of the warm Jupiter 55 Cnc b with the settings of this simulation. 

\subsection{Detectability for Warm Planets at 10 pc}

\begin{figure}[htb]
  \includegraphics[width=0.99 \linewidth]{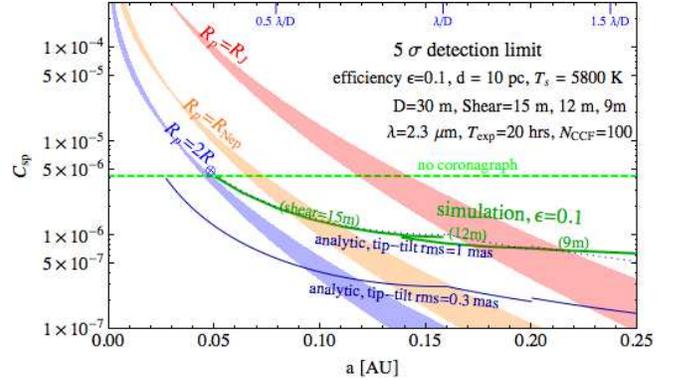}
\caption{5 $\sigma$ detection limit for planets around nearby solar-type stars for the simulated wavefront after the ExAO (green solid curve). The blue dotted curve over the green solid curve is the empirical model given by Equation (\ref{eq:estmtt}). The dashed green line indicates the expected S/N for no coronagraph  (the photon correcting area limit). The blue solid curve is the detection limit for a more ambitious tip-tilt correction (rms=0.3 mas). We choose adequate shears ($s=$15m, 12m, and 9m) according to the angular separation. The other assumptions are summarized in the right upper side of this figure. The conversion of semi-major axis $a$ to $\lambda/D$ is shown in the upper line of this panel. We assume that the efficiency of the high-dispersion instruments to be $\epsilon=$ 0.1. The red, orange, and blue shaded regions are the expected contrast assuming the equilibrium temperature with Jupiter, Neptune, and two times the Earth radii. The semimajor axes that correspond to 0.5$\lambda/D$,1$\lambda/D$, and 1.5$\lambda/D$ are shown by blue bars on the upper line of Figure. \label{fig:etametapc}}
\end{figure}

Based on the results of the simulation and Equation (\ref{eq:snest}), we also compute detectability for a hypothetical planetary system around a solar analogue at 10 pc (Figure \ref{fig:etametapc}). We assume $N_\mathrm{CCF}=100$, $T_\mathrm{exp}=20$ hr and $\epsilon=0.1$ as the efficiency. The red, orange, and blue shaded regions are the contrast computed by Equation (\ref{eq:snest}) for Jupiter, Neptune, and two times the Earth radii, respectively. As shown by the green curve in Figure \ref{fig:etametapc}, the spectroscopic coronagraph for our fiducial settings reduces the detection limit in three to six times lower than that with no coronagraph  (dashed green line) for 0.1-0.25 AU and can detect the PRV of the Jupiter-sized and Neptune-sized planets within $\sim 0.2$ and $0.1$ AU, respectively. 

To consider a more ambitious situation, we extrapolate the results to better wavefront control systems. As described above, the low-order wavefront errors also contribute the contrast as well as the tip-tilt error. The analytic expression of Equation (\ref{eq:app2ttslim}) is inaccurate. In fact, if adopting the rms of the tip-tilt $\tilde{\theta}_{t} \sim 1$ mas (i.e., $\tilde{\theta}_{t,x} \sim 0.7$ mas) provides $\sim$ 2 times larger $M$ for our simulation, as described above, where $\tilde{\theta}$ denotes the rms of $\theta$. We empirically add this factor, $K$, to estimate the gain in the S/N for the detection limit as
\begin{eqnarray} 
M(\tilde{\theta_{t}}) = K \, \sqrt{\delta (q)} \, \psi_\mathrm{lim} (\theta_p,\tilde{\theta_{t}}/\sqrt{2}), 
\label{eq:estmtt}
\end{eqnarray}
where we adopt $K = 2$. This empirical model provides a good approximation for our simulated results as shown by the dotted blue line in Figure \ref{fig:etametapc}. Scaling with Equation (\ref{eq:estmtt}), we expect a $\sim$ 10 - 30 times lower detection limit (solid blue curve in Fig [\ref{fig:etametapc}]) than that with no coronagraph if the wavefront control achieves 0.3 mas of the rms of the tip-tilt. If such ExAOs becomes reality, the spectroscopic detection for the super-Earth-sized planet $<0.1$ AU will be possible with a 30 m telescope. The possibility of further correction of the tip-tilt and the low-order aberration is discussed in Section 4.

The separations of 0.5$\lambda/D$, 1$\lambda/D$, and 1.5$\lambda/D$ are shown on the upper line of Figure \ref{fig:etametapc}. If defining the IWA of the spectroscopic coronagraphs by the angle so that the gain is conveniently twice ($M=2$), the IWA for the simulation (rms = 1 mas), and that for rms = 0.3 mas, are $\theta_p \sim $ 0.5 $\lambda/D$ and 0.25 $\lambda/D$, respectively. The IWA of the coronagraph for direct imaging is classically defined by the angle with 50 \% transmissivity, corresponding to $\theta_{1/2} = 0.25 \lambda/s$ for the two-beam nuller. The shear of the nuller used for direct imaging is set to 10 - 20 \% of the telescope diameter to obtain resolved images. The IWA of the nuller for direct imaging is typically $\theta_{1/2} \gtrsim 1 \lambda/D$. For our setting, $s=D/2$ yields $\theta_{1/2} = 0.5 \lambda/D$. The spectroscopic coronagraph achieves a small IWA by extending the shear to 30 - 50 \% of the telescope diameter.

\section{Discussion}

As described above, the suppression of the tip-tilt error  and the improvement of the low-order wavefront sensing are crucial for our technique. Though further discussion of improvement of the ExAO is beyond the scope of this paper, we briefly mention several factors that we did not consider in our simulation. One of reasons why our ExAO simulation does not achieve $< 1$ mas of the tip-tilt error is the chromatic aberration -- the use of different bands for sensing (I band) and science (K band). The wavefront sensing with a band that is close to the science band will improve the low-order aberration. The tip-tilt sensors with science lights under development will improve the situation \citep[e.g.][]{2009ApJ...693...75G,2014arXiv1404.7201S}. They proved that the sensor can reduce the tip-tilt within $\sim 0.1 \lambda/D$ in the laboratory experiment and the theoretical limitation is  $\sim 0.01 \lambda/D$ \citep{2009ApJ...693...75G}.  For instance, the stellar light rejected by the Lyot stop for the wavefront sensing can be used for the wavefront sensing with the science band.

In this paper, we have regarded the semimajor axis as the angular separation of the planets, i.e., we have assumed that the planet is located at the phase angle near $\beta \sim \pi/2$. The planets with less efficient redistribution of heat at this location have less emissivity than those at the far side of the star ($\beta \sim \pi$). This effect might reduce the benefit of the spectroscopic coronagraph. However, the PRV measurement for various phase angles is particularly important for the complementary applications of the PRV, such as the detection of the planetary spin or the zonal winds \citep[e.g.][]{kawahara12}. 


Though we have assumed the warm/hot planets close to a host star, the spectroscopic coronagraph is also applicable to the self-luminous planets. 

\section{Summary}

In this paper, we demonstrated the concept of the spectroscopic coronagraphy for future spectroscopic direct detection. We assumed the visible nuller as the spectroscopic coronagraph. As shown in the paper, the tip-tilt error and low-order wavefront control is crucial for this technique. The visible nuller with the ExAO in the next generation ELTs gain an S/N that is three to six times larger for the PRV detection of the nearby planets at $0.1-0.2$ AU. We showed that the spectra of warm Jupiters in $\lesssim 0.2$ AU and  warm Neptunes in $\lesssim 0.1 $ AU are detectable with a 30 m telescope and future ExAO. If we can utilize the more progressive ExAO with the tip-tilt rms $\sim 0.3$ mas, the improvement in the S/N is more than an order of magnitude and the spectrum of warm super-Earth-sized planets will be detectable. This paper demonstrates the importance of the tip-tilt and low-order wavefront control of ExAO on ELTs, and shows that doing so will expand the range of application of the planetary radial velocimetry.
   
We thank Masaru Kino for fruitful discussion. We also thank an anonymous referee for a lot of constructive and insightful comments. This work is supported by the Astrobiology Project of the CNSI, NINS (AB251009), and by a Grant-in-Aid for Scientific research from JSPS and from the MEXT (Nos. 25800106, 25247021,25286073, and 24103501).  

\appendix
\section{Analytical Formulation of the Visible Nuller}

In this section, we review the analytical formulation of the visible nuller and derive the transmittance. We start from a general formula including the wavefront error $\phi(\xv)$ at the pupil, which is used in the simulations in Section 3.  Given $\phi(\xv)$ at the coordinate on the pupil plane,  $\xv = {\bf X}/\lambda$ (${\bf X}$ is the physical coordinate on the pupil), the complex amplitude of the (single) pupil from the source at $\thetav$ is written as 
\begin{eqnarray}
A_{P,\mathrm{single}} (\xv,\thetav) = M_{t}(\xv) V (\thetav) e^{2 \pi i \xv \cdot \thetav + i \phi(\xv)} ,
\end{eqnarray}
where $V(\thetav)$ is the amplitude distribution on the sky plane. Quantities at pupil and focal planes are tagged by $P$ and $F$, respectively. The shape of the pupil is characterized by the mask function, $M_{t}(\xv)=1$ for the area of mirror and $=0$ for the shadows of the secondary mirror, the segment gap, the spider, and the outside area of the pupil. The combined complex amplitude of the $n$-beam nuller from $\thetav$ at the pupil is given by
\begin{eqnarray}
\label{eq:apaaa}
A_P (\xv,\thetav) =  M_o(\xv) \frac{1}{n} \sum_k^n W_k A_{P,\mathrm{single}} (\xv-\sv_k,\thetav)  = M_o(\xv) \frac{V (\thetav)}{n} \sum_k^n W_k e^{2 \pi i (\xv -\sv_k) \cdot \thetav + i \phi(\xv-\sv_k)} 
\end{eqnarray}
where $\sv_k$ is the baseline vector from the center  normalized by $\lambda$ and $W_k$ is the phase shift of the $k$-beam. Here we include the additional mask to eliminate the non-overlapped area of beams in the pupil, 
\begin{eqnarray}
M_o(\xv) = \displaystyle{ \left\{
\begin{array}{cc}
1 & \mbox{\,\,\,\,\, for $\xv \in \mathcal{S}_o (q)$,} \\
0 & \mbox{\,\,\, otherwise} \\
\end{array} \right.}
\end{eqnarray}
where $\mathcal{S}_o (q)$ is the overlapped area that the following equation satisfies:  
\begin{eqnarray}
\prod_k M_t (\xv-\sv_k) = 1,
\end{eqnarray}
and $q \equiv s/D$. Because $M_o \in M_t(\xv-\sv_k)$, we remove the term $M_t$ from Equation (\ref{eq:apaaa}).

The complex amplitude without the phase aberration at the focal plane is computed by
\begin{eqnarray}
A_F (\Thetavf, \thetav) &=& \mathbb{FT}_\xv [A_P (\xv, \thetav)] = \frac{|A_0 (\thetav)|}{n} \mathbb{FT}_\xv [M_o] \ast \int d \xv e^{- 2 \pi i \xv \cdot \Thetavf} \sum_k^n W_k e^{2 \pi i (\xv -\sv_k) \cdot \thetav} \nonumber \\
&=& \frac{|A_0 (\thetav)|}{n} \tilde{M_o} \ast [ \gamma(\thetav) \delta_D(\Thetavf - \thetav) ] = \frac{|A_0 (\thetav)|}{n} \gamma(\thetav) \tilde{M_o}(\Thetavf) \ast \delta_D(\Thetavf - \thetav) \nonumber \\
&=& \frac{|A_0 (\thetav)|}{n} \gamma(\thetav) \tilde{M_o} (\Thetavf - \thetav), 
\end{eqnarray}
where $\gamma(\thetav)$ is the amplitude transmittance,
\begin{eqnarray}
  \label{eq:gammatra}
 \gamma(\thetav) \equiv \sum_k^n W_k e^{- 2 \pi i \sv_k \cdot \thetav},
\end{eqnarray}
The visible nullers choose the adequate phase shift for $W_k$ so that the central star is canceled, i.e. $\gamma(\thetav \to {\bf 0}) \to {\bf 0}$.

The visible nullers choose the adequate phase shift for $W_k$ so that the central star is suppressed, i.e. so that the intensity of the source is $\thetav \to {\bf 0}$, 
\begin{eqnarray}
I_f (\thetav) &\propto& \Tf(\thetav) \to 0 \mbox{\,\, for $\thetav \to {\bf 0} $ },
\end{eqnarray}
where $\Tf(\thetav) = |\gamma(\thetav)|^2$ is the intensity transmittance.

The two beams nuller (the second-order nuller) has the set of $n=2$,$\sv_1 = (+s/2 \lambda,0),\sv_2 = (-s/2 \lambda,0)$ ($s$ is the separation of the shear and $\lambda$ is the wavelength ) and $W_1=1, W_2=-1$, which provides
\begin{eqnarray}
\label{eq:twobfun}
\gamma(\theta_1) = \frac{e^{- \pi i \theta_1/\theta_n}  - e^{ \pi i \theta_1/\theta_n}}{2} =  -i \sin{\left(\pi \frac{\theta_1}{\theta_n} \right)},
\end{eqnarray}
where $\theta_1$ and $\theta_2$ are the $x$ and $y$ directions of $\thetav$ and we denote the angular length of a fringe by 
\begin{eqnarray}
\theta_n \equiv \frac{\lambda}{s} \approx 32 \, \left( \frac{\lambda}{2.3 \mu \mathrm{m}} \right) \left( \frac{s}{15 \mathrm{m}} \right)^{-1} \mathrm{\,(mas)},
\end{eqnarray}
which should be regarded as  a characteristic scale of the nuller.

We denote the vector from a star to a planet by $\thetav_p$. Since the star is located on-axis, $\thetav = {\bf 0}$, $\theta_p \equiv |\thetav_p|$ describes the angular separation of the planet and the main star.

The deficit of aperture, that is, the ratio between the overlapped area of the beams and the effective area of the telescope is defined by
\begin{eqnarray}
\label{eq:tra}
\delta (q) &\equiv& \int d \xv M_o(\xv)/S_t 
\end{eqnarray}
where $q \equiv s/D$, $S_t$ is the effective area of the telescope ($S_t = \pi D^2/4$ for a circular telescope with the diameter $D$). The deficit function $\delta (q)$ for a circular pupil for the two-beam nuller is given by
\begin{eqnarray}
\label{eq:tradelq}
\delta(q) = \frac{2}{\pi} \left( \cos^{-1} q - q \sqrt{1-q^2} \right),
\end{eqnarray}
for $0 \le q \le 1$.

\end{document}